# Fractional counting of citations in research evaluation:

## A cross- and interdisciplinary assessment of the Tsinghua University in Beijing

Ping Zhou

Institute of Scientific and Technical Information of China, 15[th] Fuxing Road, Haidian District, Beijing, China. Email: zhoup@istic.ac.cn

&

Loet Leydesdorff

Amsterdam School of Communications Research (ASCoR), University of Amsterdam, Kloveniersburgwal 48, 1012 CX, Amsterdam, The Netherlands. E-mail: loet@leydesdorff.net ; http://www.leydesdorff.net

**Abstract**

In the case of the scientometric evaluation of multi- or interdisciplinary units one risks to compare apples with oranges: each paper has to be assessed in comparison to an appropriate reference set. We suggest that the set of citing papers can be considered as the relevant representation of the field of impact. In order to normalize for differences in citation behavior among fields, citations can be fractionally counted proportionately to the length of the reference lists in the citing papers. This new method enables us to compare among units with different disciplinary affiliations at the paper level and also to assess the statistical significance of differences among sets. Twenty-seven departments of the Tsinghua University in Beijing are thus compared. Among them, the Department of Chinese Language and Linguistics is upgraded from the 19[th] to the second position in the ranking. The overall impact of 19 of the 27 departments is not significantly different at the 5% level when thus normalized for different citation potentials.

**Keywords**: evaluation, interdisciplinary, department, comparison, citation, fractional

**Introduction**

When one evaluates bibliometrically a multi-disciplinary unit such as a university with different faculties, one has to normalize for systematic differences in publication and citation behavior among fields of science. For example, the number of references in a mathematics paper is systematically lower than in disciplines such as biomedicine. Systematic differences can even occur among specialties within the same discipline: journals in toxicology, for example, have impact factors significantly lower than in immunology because of lower numbers of references provided by the authors . Productivity varies among disciplines, and so do their citation cultures (Garfield, 1979, 1982; Leydesdorff, 2008).

Bibliometricians have explored various ways for solving this normalization problem. For example, a field normalized citation score (*CPP/FCSm*)[1] was first developed by the Center for Science and Technology Studies (CTWS) in Leiden (Moed *et al*., 1995). This indicator was recently modified into the mean normalized citation score (*MNCS*;[2] Waltman *et al*., 2011). *CPP/FCSm* is also known as the "Crown Indicator" of the Leiden unit. The Center for R&D Monitoring (ECOOM) at Louvain designed a similar indicator, the normalized mean citation rate *NMCR* (Glänzel *et al*., 2009; Schubert & Braun, 1986).

---

[1] The abbreviation "*CPP*" is used by CWTS for the average of the number of citations per publication, and *FCSm* for the mean field-normalized citation score.
[2] "*MNCS*" is used by CWTS as an abbreviation of "mean normalized citation score."

The "crown indicator" *CPP/FCSm* and this latter indicator divide the average number of citations to a document set to be evaluated by the average number of citations in the relevant field of science. Fields of science are then operationalized in terms of journal sets.

Opthof & Leydesdorff (2010) argued that such a quotient of means can no longer be used for statistical testing or the indication of error of the measurement. In their opinion, one should first normalize for each individual paper against a reference set and only average thereafter over this distribution (Gingras & Larivière, 2011; Lundberg, 2007). This new indicator is called by CWTS their "new crown indicator" or *MNCS*, and in the meantime also applied in the Leiden Rankings (2010) of universities. The new indicator has the advantage of being mathematically consistent while the previous "crown indicator" was not (Waltman *et al.*, 2011).

A remaining problem is posed by the assumptions involved in the delineation of fields. Field delineation is needed for the normalization. Fields at the (sub)disciplinary level are often defined in terms of the 222 ISI Subject Categories which are attributed to the journals included in the *Journal Citations Reports* of the (*Social*) *Science Citation Index*.[3] However, these categories were designed for the purpose of information retrieval and not for the scientometric evaluation. The subject categories lack an analytical base (Pudovkin

---

[3] The *Science Citation Index* contains currently 175 Subject Categories of which three are dormant; the *Social Science Citation Index* 57 Subject Categories of which 7 overlap with the *Science Citation Index.*

& Garfield, 2002, at p. 1113n.; Rafols & Leydesdorff, 2009; Leydesdorff & Opthof, 2011) and have the following flaws: 1) journals involving more than a single field can be assigned to up to five different fields, 2) journals in the same subject category may have different disciplinary affiliations, and 3) even articles in the same journals, especially the multidisciplinary ones, may belong to different fields of science.

In summary, field classifications based on groups of journals (such as the ISI Subject Categories) cannot clearly define fields in terms of articles (Leydesdorff, 2006; Ball *et al*., 2009). Zitt *et al.,* 2005, at p. 391), however, formulated the urgency of clarity about the field-normalization in scientometric evaluations as follows: "(F)ield-normalized indicators are not only, trivially, dependent on the delineation of fields, but also, for a given multi-level classification, dependent on the hierarchical level of observation in a particular classification. An article may exhibit very different citation scores or rankings when compared within a narrow specialty or a large academic discipline."

In order to avoid the problems of defining fields in terms of the set of ISI Subject Categories—or another categorization of journals (e.g., Glänzel & Schubert, 2003; Rafols & Leydesdorff, 2009)—Ball *et al*. (2009), for example, proposed the *J* indicator. This indicator compares the citation rates of a unit under study with the weighted citation rates of each of the journals in which this unit has published. Consequently, the indicator is comparable to the the mean Journal Citation Score of Leiden (*JCSm*), but unlike

*CPP/JCSm*, the resulting *J* is a statistic based on the average of ratios and not a ratio between averages (Gingras & Larivière, 2011).[4]

In the calculation of the *J* factor, each citation is treated equally without differentiating among the weights of citations. However, articles in a reference list of a citing article are published in journals of variable prestige. Those published in a more prestigious journal can also be weighted higher than those in less prestigious ones (Pinski & Narin 1976; Cronin, 1984; Davis, 2008). In addition to the number of citations that a journal receives from other journals, the status of a journal can also be decided by the prestige of the citing journals (Franceschet, 2010). Timeliness of citations (e.g., citation half-lives) can also be considered as playing a role in deciding upon an article's prestige (Walker, et al., 2007; Sayyadi & Getoor, 2009; Järvelin & Persson, 2008; Yan & Ding, 2010).

In summary, many elements may play a role in deciding on an article's or a journal's citation impact. In another contribution to this discussion, Leydesdorff and Opthof (2010a) suggested that the definition of fields in terms of journals has hitherto not proved to be a fruitful heuristics. Journals are themselves mixed bags and increasingly so because the internet facilitates reading of papers across journals. The reduction of fields to sets of relevant journals can, in the opinion of these authors, be replaced by a definition of the relevant fields of impact in terms of the citing papers.

---

[4] The *J* indicator can be compared with the journal-equivalent of *MNCS* in the new scheme of CWTS; this is: *MNCS/MNJS* (Van Raan *et al.*, 2010, at p. 291). At this moment, however, it is still unclear whether or not this is another quotient between two means.

The reasoning behind the normalization using fractional counting is as follows: if differences among fields are caused by differences in citation behavior among citing authors (with different disciplinary identities), then normalization should be in terms of the sources of these differences, that is, at the level of individual (citing) papers. Fractional counting of citations provides a means to control for the in-between field differences caused by different citation potentials (Garfield, 1982; Moed, 2010). For example, a citation in a reference list of six items (like in mathematics) can be weighted as 1/6 in the overall citation count, while a citation among 40 references then counts as 1/40. Remaining differences may be caused by the different citation half-lives among disciplines and thus be attenuated by using a longer citation window.

The method of fractional counting was first proposed by Price & DeBeaver (1966) for the proportionate attribution of co-authorships to papers, and has since been used more extensively in research evaluations (e.g., National Science Board, 2010; cf. Narin, 1976). Fractional counting may lead to a perspective very different from integer counting when applied to addresses (Anderson *et al*., 1988; Leydesdorff, 1988). In this study, however, we use fractional attribution of the references in the citing documents to the cited documents, while in case of fractional counting among authors or addresses the cited documents are fractionated in the publication analysis.

One can also combine the two effects and study their interactions using different weighting schemes (Galam, 2010; Neufeld & Von Iens, in press), but in this study we focus on citation analysis and fractionate only in terms of the citing papers (Leydesdorff & Shin, accepted). For analytical reasons, we compare the fractional counts directly with the integer counted citations because mixing the effect of this change of perspective with another form of field-normalization based on journal groupings (as sometimes advocated; Moed, 2010; Zitt, 2010) can make it difficult for the reader to follow the effects of citing-side normalization.

Fractional counting was first applied to citation analysis by Small & Sweeney (1985) for generating co-citation maps and also used by Zitt & Small (2008) for journal normalization. Moed (2010) suggested this method for the normalization when developing the Source-Normalized Impact per Paper (SNIP) indicator of *Scopus*, but the SNIP indicator is again a composed quotient of an average divided by a median, and thus not a proper statistic (Leydesdorff & Opthof, 2010b; Moed, 2011). In other words, the coverage of the Scopus database was assumed as the system of reference when developing the SNIP indicator, and not the set of citing articles.

Fractional counting of citations was first applied to research evaluation by Leydesdorff & Opthof (2010a and b), but on a limited set of documents. Leydesdorff and Bornmann (in press) scaled this method up to the recalculation of the impact factors in the journal set of the *Science Citation Index*. Using the thirteen fields identified by ipIQ for the purpose of developing the Science and Engineering Indicators 2010 (NSB, 2010, at p. 5-30 and Appendix Table 5-24), it could be shown that normalization by fractional counting reduces the in-between group variance in the impact factors (2008) by 81% (when compared with integer counting) and made the remaining differences statistically not significant.

In this study, we apply fractional counting of the citations for the first time to a large set for an institutional evaluation, namely, the departments of one of the leading universities of China, the Tsinghua University. We distinguish among 27 departments in different disciplines and show that this correction for the in-between field variation changes the rank order in important respects.

## 2. Data and methods

Data were retrieved from the online version of the Web of Science (WoS) of Thomson Reuters. Sources included the *Science Citation Index Expanded* (*SCI-EXPANDED*), the *Social Sciences Citation Index* (*SSCI*), the *Conference Proceedings Citation Index - Science* (*CPCI-S*), and the *Conference Proceedings Citation Index - Social Science & Humanities* (*CPCI-SSH*).[5] Only articles, reviews and proceeding papers ($N = 3,950$) published in 2005 ("py = 2005") were included. The publication year 2005 was chosen in order to be able to use a five-year window. These 3,950 documents are the documents to be evaluated in this study.

---

[5] The data was collected online at the WoS interface of the Institute of Scientific and Technical Information of China (ISTIC) which does not provide access to the *A&HCI*. At the University of Amsterdam (UvA) one has access to the *A&HCI* but not to the *CPCI-S* and the *CPCI-SSH*. We did retrieve data from the *A&HCI* at this end. Three of the four publications thus retrieved did not provide departmental information and none of them were cited since their publication in 2005. The single record with departmental information was from the Center of Library Education which was not defined as a unit under study. In summary, whether or not the *A&HCI* data is covered cannot be expected to affect the results other than marginally.

The WoS interface conveniently allows downloading the citing documents of these 3,950 cited documents. We use—for reasons to be explained below—a five-year time window (2005-2009). Each of the (16,882) citing documents contains a number of references ($k$) and is accordingly attributed to the cited document with a fractional weight $1/k$. Cited and citing documents are aggregated in terms of the departmental structures of the Tsinghua University in Beijing and then compared both in terms of numbers (of citations and publications) and in terms of their respective impacts (citations/publications).

Various problems have to be addressed when defining the different units to be evaluated. Tsinghua University is organized at three levels: (*i*) schools (or colleges), (*ii*) departments, and (*iii*) laboratories or research centers. Most schools are composed of departments, but this is not always the case. For example, the School of the Life Sciences, the School of Public Policy & Management, and the School of Law do not contain subsidiary organizations like departments. Some departments (such as the Department of Environmental Science & Engineering, the Department of Electrical Engineering) are not affiliated with schools. Different departments collaborating in the same school usually have other disciplinary affiliations. Research centers or laboratories are in most cases under the management of a department, but some of them are affiliated with more than a single department. For example, the Key Laboratory of Atomic and Molecular

Nanosciences of the Ministry of Education is a joint enterprise of the three departments of Physics, Chemistry, and Engineering Physics.

In addition to the complexity of the organizational affiliations, the ways in which authors mark their affiliations in publications vary. Some authors provide school names and others department names. This makes it difficult to assign publications unambiguously. Taking the above issues into consideration, we decided to consider schools or colleges with no departmental affiliation in addition to departments as units of analysis. Problems encountered in the assignment of publications to affiliations could thus be avoided. All departments or schools with no departmental affiliations will be called "departments" in the remainder of this study.

Thus organized, 64 departments were distinguished in Tsinghua University. Since this study is based on scholarly publications and statistical in nature, departments with less than five publications were not included. This left us with 27 departments amenable to the bibliometric evaluation. Of the publications published by Tsinghua University in 2005, 4,766 were indexed in the Web of Science at the date of retrieval (August 2010). The 27 departments evaluated in this study published 82.9% of the publications of Tsinghua University covered by the WoS in 2005. As both Mainland China and Taiwan have universities with "Tsinghua" in their names and the current study focuses on the university in the Mainland, we excluded Taiwan in the queries.[6]

---

[6] A query for publication data of the Department of Physics reads as follows:

(continued)

Although "Tsinghua" is the official name for the university, individual authors might wish to use the Chinese Pinyin "Qinghua" to refer to the university. After checking in the Web of Science using "Qinghua Univ" or "Qing Hua Univ" for publications in 2005, however, we found only one single record deviating. Thus, spelling variations can be ignored among the publications of this university.

First, publication data of each department were retrieved. Thereafter, data of the citing papers were harvested and related to the cited set using dedicated routines. SPSS was used for the statistical analysis. The 27 departments can be compared in terms of integer

---

ad=(tsing hua univ same sch Phys or tsing hua univ same phys sch or tsing hua univ same Dep phys or tsing hua univ same phys Dep or tsing hua univ same coll phys or tsing hua univ same phys coll or tsinghua univ same Dep phys or tsinghua univ same phys Dep or tsinghua univ same sch phys or tsinghua univ same phys sch or tsinghua univ same coll phys or tsinghua univ same phys coll) and ad=(china not taiwan) and py=2005.

For publications of some departments one may need more than a single query. For example, for the retrieval of the publications of the Department of Chemistry one needs three queries. The first query covers publications of both the Department of Chemistry and the Department of Chemical Engineering:

1) ad=(tsing hua univ same sch Chem or tsing hua univ same chem sch or tsing hua univ same Dep chem or tsing hua univ same chem Dep or tsing hua univ same coll chem or tsing hua univ same chem coll or tsinghua univ same Dep chem or tsinghua univ same chem Dep or tsinghua univ same sch chem or tsinghua univ same chem sch or tsinghua univ same coll chem or tsinghua univ same chem coll) and ad=(china not taiwan) and py=2005

2) ad=(tsing hua univ same sch Chem eng or tsing hua univ same chem eng sch or tsing hua univ same Dep chem eng or tsing hua univ same chem eng Dep or tsing hua univ same coll chem eng tsing hua univ same chem eng coll or tsinghua univ same Dep chem eng or tsinghua univ same chem eng Dep or tsinghua univ same sch chem eng or tsinghua univ same chem eng sch or tsinghua univ same coll chem eng or tsinghua univ same chem eng coll) and ad=(china not taiwan) and py=2005

3) Results of 1) – results of 2) by using "Not": #1 not #2

counted citations using standard measures (such as the c/p ratios), but the distribution of fractional counts allows us to run some additional statistics.

For this purpose, the 27 departments can be considered as independent samples of unequal size and Kruskall-Wallis then provides an appropriate test. Using ANOVA, one is additionally able to compare the units under evaluation among them and determine whether they are significantly different in terms of the citation distributions using a *post-hoc* test. Among these tests, we use Dunnett's C-test because the variance among the samples was not homogeneous (Levene's test).

In order to enhance the readability of the results of comparing the 27 departments (that is, (27 x 26)/2 = 351 combinations), one can visualize the similarities in this respect among these 27 departments as a network in which homogenous groups are considered as components of a graph which are linked together. (The density of this network provides us additionally with a global measure of equality among the departments; cf. Leydesdorff & Bornmann, in press). Departments which are not linked to each other by an edge of the network are significantly different in their impact at the level of $p < 0.05$.

**3. Results**

Publication and citation parameters of the 27 departments are provided in Table 1. Since the time between publication and citation also varies among fields of science, we first compared citation counts using two citation windows: three years (2005-2007) and five years (2005-2009) given that 2005 was the year of publication. Table 1 shows that the departments can be ordered differently based on the various parameters.

**Table 1.** Different counting scores of departments of Tsinghua University;
P= Number of Publications; IC= Integral Counts of Citations; FC= Fractional Counts of Citations.

| Department | P (2005) | Three-year citation window (2005-2007) | | | | Five-year citation window (2005-2009) | | | |
|---|---|---|---|---|---|---|---|---|---|
| | | IC | IC/P | FC | FC/P | IC | IC/P | FC | FC/P |
| Dep Automat | 270 | 374 | 1.39 | 22.13 | 0.08 | 906 | 3.36 | 47.46 | 0.18 |
| Dep Automot | 5 | 3 | 0.6 | 0.16 | 0.03 | 8 | 1.6 | 0.3 | 0.06 |
| Dep Biomed Engn | 91 | 108 | 1.19 | 5.16 | 0.06 | 246 | 2.7 | 10.02 | 0.11 |
| Dep Bldg Sci | 46 | 46 | 1 | 2.48 | 0.05 | 111 | 2.41 | 5.6 | 0.12 |
| Dep Chem | 404 | 2080 | 5.15 | 73.91 | 0.18 | 4950 | 12.25 | 166.36 | 0.41 |
| Dep Chem Engn | 191 | 506 | 2.65 | 19.96 | 0.1 | 1146 | 6 | 41.98 | 0.22 |
| Dep Chinese Languages | 5 | 9 | 1.8 | 1.17 | 0.23 | 11 | 2.2 | 1.31 | 0.26 |
| Dep Civil Engn | 35 | 53 | 1.51 | 2.73 | 0.08 | 138 | 3.94 | 6.9 | 0.2 |
| Dep Comp Sci & Tech | 392 | 250 | 0.64 | 14.4 | 0.04 | 542 | 1.38 | 30.14 | 0.08 |
| Dep Econ | 43 | 42 | 0.98 | 1.85 | 0.04 | 103 | 2.4 | 4.71 | 0.11 |
| Dep Elect Engn | 507 | 379 | 0.75 | 30.84 | 0.06 | 795 | 1.57 | 58.71 | 0.12 |
| Dep Engn Mech | 185 | 422 | 2.28 | 19.93 | 0.11 | 882 | 4.77 | 39.02 | 0.21 |
| Dep Engn Phys | 81 | 95 | 1.17 | 7.51 | 0.09 | 156 | 1.93 | 11.02 | 0.14 |
| Dep Environm Sci & Engn | 76 | 214 | 2.82 | 8.03 | 0.11 | 548 | 7.21 | 18.39 | 0.24 |
| Dep Ind Engn | 22 | 21 | 0.95 | 0.75 | 0.03 | 39 | 1.77 | 1.5 | 0.07 |
| Dep Mat Sci | 7 | 17 | 2.43 | 0.92 | 0.13 | 27 | 3.86 | 1.35 | 0.19 |
| Dep Mat Sci & Engn | 543 | 1037 | 1.91 | 49.66 | 0.09 | 2366 | 4.36 | 105.46 | 0.19 |
| Dep Mech Engn | 145 | 237 | 1.63 | 11.03 | 0.08 | 546 | 3.77 | 24.75 | 0.17 |
| Dep Pharmaceut Sci | 12 | 30 | 2.5 | 1.32 | 0.11 | 64 | 5.33 | 2.73 | 0.23 |
| Dep Phys | 305 | 1062 | 3.48 | 43.65 | 0.14 | 2032 | 6.66 | 79.04 | 0.26 |
| Dep Precis & Mechanol | 221 | 266 | 1.2 | 19.36 | 0.09 | 523 | 2.37 | 34.37 | 0.16 |
| Dep Thermal Engn | 86 | 70 | 0.81 | 3.2 | 0.04 | 183 | 2.13 | 8.01 | 0.09 |
| Inst Microelect | 88 | 59 | 0.67 | 3.84 | 0.04 | 151 | 1.72 | 6.9 | 0.08 |
| Inst Nucl & New Energy Tech | 82 | 87 | 1.06 | 4.81 | 0.06 | 194 | 2.37 | 9.62 | 0.12 |
| Sch Life Sci | 32 | 67 | 2.09 | 2.24 | 0.07 | 149 | 4.66 | 5.43 | 0.17 |
| Sch Publ Policy & Management | 12 | 17 | 1.42 | 1.09 | 0.09 | 35 | 2.92 | 2.02 | 0.17 |
| Sch Software | 64 | 25 | 0.39 | 1.31 | 0.02 | 67 | 1.05 | 3.28 | 0.05 |

The departments of Materials Science & Engineering, Electrical Engineering, Chemistry, and Computer Science and Technology lead the ranking in terms of numbers of publications. In terms of citations based on integer counting, the rank order changes: only the Departments of Chemistry and Materials Science & Engineering are still part of the top four. The rank order would again be different using different citation windows or the method of fractional counting of the citations.

**3.1 Two types of parameters: aggregated citations and c/p ratios**

As discussed above, methods based on integral counting cannot be used for cross-disciplinary assessments without normalization. Fractional counting normalizes disciplinary variations in terms of the length of reference lists in scholarly literature. Using SPSS, we investigated correlations between the two counting methods (Table 2). The (Spearman) rank-order correlations are provided in the upper triangle, and the Pearson moment correlations in the lower one.

**Table 2**. Pearson and (Spearman) rank correlations between different parameters in the lower and upper triangle, respectively ($N = 27$).

|  | P (2005) | IC/P (05-07) | IC/P (05-09) | FC/P (05-07) | FC/P (05-09) | IC (05-07) | IC (05-09) | FC (05-07) | FC (05-09) |
|---|---|---|---|---|---|---|---|---|---|
| **P (2005)** |  | 0.093 | 0.133 | 0.093 | 0.12 | .934(**) | .927(**) | .946(**) | .954(**) |
| **IC/P (05-07)** | 0.248 |  | .942(**) | .890(**) | .941(**) | .386(*) | .385(*) | 0.347 | 0.324 |
| **IC/P (05-09)** | 0.281 | .967(**) |  | .729(**) | .847(**) | .422(*) | .440(*) | 0.369 | 0.362 |
| **FC/P (05-07)** | 0.111 | .744(**) | .598(**) |  | .945(**) | 0.328 | 0.303 | 0.349 | 0.301 |
| **FC/P (05-09)** | 0.259 | .936(**) | .890(**) | .872(**) |  | .382(*) | 0.38 | .393(*) | 0.352 |
| **IC (05-07)** | .715(**) | .756(**) | .783(**) | .465(*) | .698(**) |  | .988(**) | .983(**) | .988(**) |
| **IC (05-09)** | .698(**) | .753(**) | .792(**) | .457(*) | .700(**) | .996(**) |  | .977(**) | .983(**) |
| **FC (05-07)** | .843(**) | .652(**) | .673(**) | .415(*) | .625(**) | .972(**) | .960(**) |  | .994(**) |
| **FC (05-09)** | .823(**) | .666(**) | .701(**) | .411(*) | .638(**) | .980(**) | .978(**) | .995(**) |  |

\*\* Correlation is significant at the 0.01 level (2-tailed).
\* Correlation is significant at the 0.05 level (2-tailed).

In addition to the effects on the ranking (using rank-order correlation), we use the Pearson correlations for the relations between the parameters (which are interval scaled).

The parameters measuring citation impact can be classified into two types: the aggregate number of citations (volume) and normalized as *c/p* ratios (impact), respectively. Volume parameters (*IC* for integral citations and *FC* for fractionated citations) depend on the size of the document set (*P*) and are strongly correlated among them. As the first row of Table 2 shows, numbers of publications and citations are strongly correlated, whereas impact is not correlated with the size of the document set.

The values for *IC* and *FC* are always correlated among them with (Pearson) correlations higher than 0.95 ($p < 0.01$); this is independent of the citation windows. The normalized impact parameters (*IC/P* and *FC/P*) are also significantly correlated among themselves, but at lower levels. Interesting are the off-diagonal comparisons (in terms of the boxes in Table 2) that show that the rank correlations between size and impact indicators are never significant at the 0.01-level and in many cases not even at the 0.05-level. This uncoupling between the size-driven total number of citations and citation impact (*c/p*) is stronger for fractional than integer counting. The two dimensions of citation analysis are thus more clearly separated using fractional counting.

The effects of the choice of a three-year or five-year citation windows lead to rank-order correlations of 0.942 in the case of integer counting (*IC/P*) and 0.945 in the case of fractional counting (*FC/P*); both correlations are significant at the 0.01-level. Thus, the two time-windows are not indicated as significantly different in terms of the rankings. The Pearson correlation between the two time windows, however, declines to 0.872 for fractional counting, while it remains high for integer counting (0.967). Fractional counting of the impact thus is somewhat more sensitive to the choice of the citation window than integer counting. This accords with Ludo Waltman's suggestion (personal communication, 23 June 2010) that the remaining differences between fields after correction for the citation potentials (by fractional counting/paper), could be caused by the different rates at which papers in the past years are cited in various fields of science (cf. Leydesdorff & Bornmann, in press).

**3.2 Ranking results using integer or fractional counting**

The citation impacts can be compared in terms of both the total impact of each department ($\Sigma c$ in Table 3) and its normalized impact ($c/p$ ratios in Table 4). As noted, a five-year citation window is used.

**Table 3.** Ranking order using total IC and FC.

| IC | Rank | FC | Rank Change |
|---|---|---|---|
| Dep Chem | 1 | Dep Chem | |
| Dep Mat Sci & Engn | 2 | Dep Mat Sci & Engn | |
| Dep Phys | 3 | Dep Phys | |
| Dep Chem Engn | 4 | Dep Elect Engn | +3 |
| Dep Automat | 5 | Dep Automat | |
| Dep Engn Mech | 6 | Dep Chem Engn | -2 |
| Dep Elect Engn | 7 | Dep Engn Mech | -1 |
| Dep Environm Sci & Engn | 8 | Dep Precis & Mechanol | +3 |
| Dep Mech Engn | 9 | Dep Comp Sci & Tech | +1 |
| Dep Comp Sci & Tech | 10 | Dep Mech Engn | -1 |
| Dep Precis & Mechanol | 11 | Dep Environm Sci & Engn | -3 |
| Dep Biomed Engn | 12 | Dep Engn Phys | +3 |
| Inst Nucl & New Energy Tech | 13 | Dep Biomed Engn | -1 |
| Dep Thermal Engn | 14 | Inst Nucl & New Energy Tech | -1 |
| Dep Engn Phys | 15 | Dep Thermal Engn | -1 |
| Inst Microelect | 16 | Dep Civil Engn | +2 |
| Sch Life Sci | 17 | Inst Microelect | +2 |
| Dep Civil Engn | 18 | Dep Bldg Sci | +1 |
| Dep Bldg Sci | 19 | Sch Life Sci | -2 |
| Dep Econ | 20 | Dep Econ | |
| Sch Software | 21 | Sch Software | |
| Dep Pharmaceut Sci | 22 | Dep Pharmaceut Sci | |
| Dep Ind Engn | 23 | Sch Publ Policy & Management | +1 |
| Sch Publ Policy & Management | 24 | Dep Ind Engn | -1 |
| Dep Mat Sci | 25 | Dep Mat Sci | |
| Dep Chinese Language & Literature | 26 | Dep Chinese Language & Literature | |
| Dep Automot | 27 | Dep Automot | |

When assessment is focused on the aggregated impact (Table 3), the rankings of 17 of the 27 departments remain unchanged. In the other ten cases, the differences in the ranks are at most three positions. As noted, the rank-order correlation ($\rho$) was 0.983 ($p < 0.01$) in this case. Yet, at the level of some individual departments, these changes may be important. For example, the Department of Precision & Mechanology and the Department of Environmental Science & Engineering swapped the $8^{th}$ and $11^{th}$ positions in the two rankings. However, the differences are moderated because both *IC* and *FC* are driven by scale effects of the respective document sets as measured in terms of the number of publications.

**Table 4.** Ranking results using IC/P and FC/P.

| IC/P | Rank | FC/P | Rank Change |
|---|---|---|---|
| Dep Chem | 1 | Dep Chem | |
| Dep Environm Sci & Engn | 2 | Dep Chinese Language & Literature | +17 |
| Dep Phys | 3 | Dep Phys | |
| Dep Chem Engn | 4 | Dep Environm Sci & Engn | -2 |
| Dep Pharmaceut Sci | 5 | Dep Pharmaceut Sci | |
| Dep Engn Mech | 6 | Dep Chem Engn | -2 |
| Sch Life Sci | 7 | Dep Engn Mech | -1 |
| Dep Mat Sci & Engn | 8 | Dep Civil Engn | +1 |
| Dep Civil Engn | 9 | Dep Mat Sci & Engn | -1 |
| Dep Mat Sci | 10 | Dep Mat Sci | |
| Dep Mech Engn | 11 | Dep Automat | +1 |
| Dep Automat | 12 | Dep Mech Engn | -1 |
| Sch Publ Policy & Management | 13 | Sch Life Sci | -6 |
| Dep Biomed Engn | 14 | Sch Publ Policy & Management | -1 |
| Dep Bldg Sci | 15 | Dep Precis & Mechanol | +2 |
| Dep Econ | 16 | Dep Engn Phys | +5 |
| Dep Precis & Mechanol | 17 | Dep Bldg Sci | -2 |
| Inst Nucl & New Energy Tech | 18 | Inst Nucl & New Energy Tech | |
| Dep Chinese Language & Literature | 19 | Dep Elect Engn | +6 |
| Dep Thermal Engn | 20 | Dep Biomed Engn | -6 |
| Dep Engn Phys | 21 | Dep Econ | -5 |
| Dep Ind Engn | 22 | Dep Thermal Engn | -2 |
| Inst Microelect | 23 | Inst Microelect | |
| Dep Automot | 24 | Dep Comp Sci & Tech | +2 |
| Dep Elect Engn | 25 | Dep Ind Engn | -3 |
| Dep Comp Sci & Tech | 26 | Dep Automot | -2 |
| Sch Software | 27 | Sch Software | |

When the impact is normalized for the size of the document set (Table 4), the rank-order correlation ($\rho$) declines to 0.847 ($p < 0.01$). Only seven of the departments are unaffected by the changes. Despite the still relatively high correlation, however, the differences are in some cases dramatic. For example, the Department of Chinese Language and Literature which was ranked 19[th] when using integer counting, obtains the second position after this correction for the between-field variation. The ranks of the Department of Engineering Physics (+5) and the Department of Electronic Engineering (+6) increase considerably, while the impacts of the School of the Life Sciences (-6) and the Departments of Biomedical Engineering (-6), and Economics (-5) are rated much lower.

In other words, publications in some engineering fields (in the natural sciences) are severely under-estimated in terms of their citation impact when one disregards the citation density among publications in these fields, while they are overestimated in the case of the biomedical sciences and engineering. The effects for economics are also considerable. When *FC* were used for the assessment of performance, however, 63% of the departments and schools would not have reasons to complain since their respective ranks would remain unaffected or increase in comparison with the results of an evaluation using integer counting. Furthermore, the decreases are less dramatic than some of the increases. Particularly, the upward evaluation of the impact of the five papers of the Department for Chinese Language and Literature rightly appreciates the different nature of publications and citations in the humanities (Garfield, 1982; Nederhof, 2006).

**3.3 Are differences also statistically significant?**

Different from impact analysis using integer counting, fractional counting of the citation impact provides us with a distribution of fractional values which allows for the testing of differences in terms of their statistical significance. The 27 document sets can be considered as independent samples of unequal size which have been cited to variable extents. The appropriate test for such a design is provided by Kruskal-Wallis and leads to a value of $\chi^2(df=26) = 1977.917; p < 0.01$. In other words, the 27 departments are significantly different in terms of their citation impact.

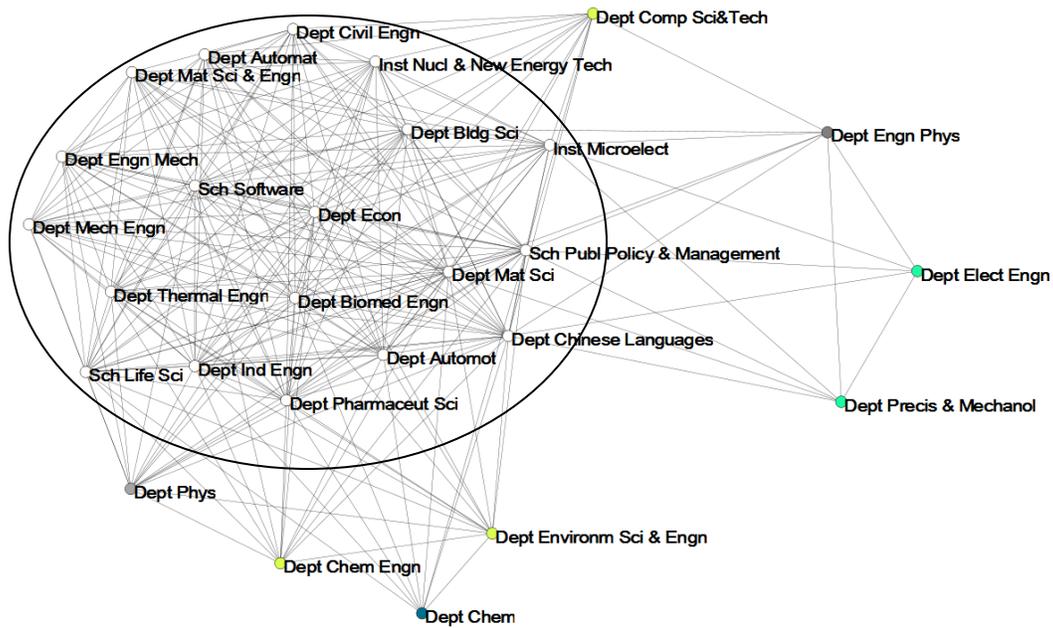

**Figure 1.** Homogeneity among the 27 departments of Tsinghua University (Beijing) in terms of their fractional citation impacts.

Additionally, ex-post correction of the comparison using Bonferroni correction (within ANOVA) allows us to test multiple comparisons on their significance. Because the variance is not homogenous in this case (the result of Levene's test is significant), we should use a derivative of the Bonferroni correction such as Dunnett's C-test.

Figure 1 provides a graphical representation (and summary) of the result of the (27 x 26) = 702 possible comparisons. An edge in this graph indicates that the two departments at the vertices are *not* significantly different in terms of their citation impact. Thus, 19 of the 27 departments are each not significantly different from one another: they form a (*k*=17) core set. The other eight departments are organized in two groups of four: one group with the departments of Physics, Chemistry, Chemical Engineering, and Environmental Science & Engineering, and a second one of a group of mainly engineering departments.

**4. Discussion and conclusions**

Indicators for measuring citation impact can be classified into two types: aggregate and average impact, respectively. Both integer counts and fractional counts (that is, *IC* and *FC*) are related to size as indicators measuring aggregate impact. Size indicators normalized by the number of publications (that is, *IC/P* and *FC/P*) measure citation impact first at the level of each individual paper. In the case of integer counting the citations are considered to be equal; fractional counting enables us to normalize in terms of the citing papers. The collection of citing papers can be considered as a representation of the relevant scientific field of the cited paper. Thus, fractional counting normalizes for differences among fields without using an a priori classification scheme of journals in terms of fields.

Fractional counting of citations solves the hitherto unsolved problem of field-normalization of citations. Previous attempts to distinguish fields in terms of journal classifications (e.g., ISI Subject Categories) failed because the aggregated journal-journal matrix cannot be fully decomposed. Hierarchical structures span over different specialties, which are also sometimes interwoven in terms of substances and methods. A journal in econometrics, for example, is part of the economics field, but also one of the journals in a "metrics" field. An unambiguous classification of articles in such journals is impossible and different weighting schemes may lead to very different ratings in the evaluation.

In this study, we applied the method of fractional counting for the first time to a large multidisciplinary unit such as a major university; in this case, the Tsinghua Unversity in Beijing. This university is well known both for its contributions to the natural sciences and engineering, and some social sciences such as economics. These two very different disciplinary structures are organized in the *Science Citation Index* and *Social Science Citation Index*, respectively. Furthermore, the Department of Chinese Language & Literature publishes in journals which are part of the *Arts & Humanities Citation Index* which is very differently organized in terms of journal categories (Leydesdorff & Salag, 2010).

Our results showed that this latter department, notably, is upgraded in its status using fractional counting for the evaluation. The other differences are more modest, but sometimes noteworthy. A large number of departments, however, have an impact that is not significantly different from each other in terms of the statistics although the set is significantly not homogeneous (using the Kruskal-Wallis test for $\chi^2$).

This study thus suggests that fractional counting improves on the cross-disciplinary assessment. However, fractional counting is not a perfect solution! An important topic for further research remains which document types play a role in the evaluation. For example, review papers often contain long lists of references, and fractional citation counts of literatures cited in this type of documents would therefore be marginalized. Furthermore, differences in citation half-lives among both disciplines and document types (Leydesdorff, 2008, at pp. 280f.) cannot properly be captured by using a citation window for the static comparison.

**Acknowledgement**

This research was supported by the National Natural Science Foundation of China (NSFC), grant number 71073153.